\documentclass[aps,preprint,superscriptaddress]{revtex4-2}
\usepackage{graphicx}
\usepackage{txfonts}
\usepackage{bm}
\usepackage{lineno}
\usepackage{color}

\begin{document}
\title{Polarization-independent second-order photonic topological corner states}



\author{Linlin Lei}
 \affiliation{School of Physics and Materials Science, Nanchang University, Nanchang 330031, China}
\author{Shuyuan Xiao}
 \affiliation{Institute for Advanced Study, Nanchang University, Nanchang 330031, China}
 \affiliation{Jiangxi Key Laboratory for Microscale Interdisciplinary Study, Nanchang University, Nanchang 330031, China}
\author{Wenxing Liu}
\affiliation{School of Physics and Materials Science, Nanchang University, Nanchang 330031, China}
\author{Qinghua Liao}
 \email{lqhua@ncu.edu.cn}
\affiliation{School of Physics and Materials Science, Nanchang University, Nanchang 330031, China}
\author{Lingjuan He}
\affiliation{School of Physics and Materials Science, Nanchang University, Nanchang 330031, China}
\author{Tianbao Yu}
 \email{yutianbao@ncu.edu.cn}
\affiliation{School of Physics and Materials Science, Nanchang University, Nanchang 330031, China}



\begin{abstract}
Recently, much attention has been paid to second-order photonic topological insulators (SPTIs), because of their support for highly localized corner states with excellent robustness. SPTIs have been implemented in either transverse magnetic (TM) or transverse electric (TE) polarizations in two-dimensional (2D) photonic crystals (PCs), and the resultant topological corner states are polarization-dependent, which limits their application in polarization-independent optics. However, to achieve polarization-independent corner states is not easy, since they are usually in-gap and the exact location in the topological bandgap is not known in advance. Here, we report on a SPTI based on a 2D square-lattice PC made of an elliptic metamaterial, and whether the bandgap is topological or trivial depends on the choice of the unit cell. It is found that locations of topological bandgaps of TM and TE polarizations in the frequency spectrum can be independently controlled by the out-of-plane permittivity $\varepsilon_\perp$ and in-plane permittivity $\varepsilon_{\varparallel}$, respectively, and more importantly, the location of in-gap corner states can also be separately manipulated by them. From this, we achieve topological corner states for both TM and TE polarizations with the same frequency in the PC by adjusting $\varepsilon_\perp$ and $\varepsilon_\varparallel$, and their robustness against disorders and defects are numerically demonstrated. The proposed SPTI provides a potential application scenario for polarization-independent topological photonic devices. 
\end{abstract}


\maketitle


\section{introduction}
Recently, the concept of higher-order topological insulators (HOTIs) has been extended from electronic waves into classic waves\cite{1,2,3,4,5,6,7,8,9,10,11,12}. It has been shown that HOTIs do not obey the usual bulk-edge correspondence but comply with the bulk-edge-corner correspondence\cite{13,131,132}. For instance, a two-dimensional (2D) second-order topological insulator possesses one-dimensional (1D) gapped edge states and zero-dimensional (0D) in-gap corner states. In addition to the characteristics of strong field localization and small mode volume, 0D corner states also show excellent robustness against fabrication flaws\cite{14,15,16}. On this basis, they have enormous application value in the topological cavity\cite{16,161}, lasing\cite{17,171}, non-linear optics\cite{172,173}, and sensing\cite{18}. However, for photonic crystals (PCs), the two kinds of polarization, transverse magnetic (TM) and transverse electric (TE) modes, are usually studied in a separate way. One reason is either of the two modes can be excited independently, each with its own band structure, and the other is that forming a common band gap (CBG) is not easy, especially the topological one. Past researches have shown the polarization-independent optics is potentially useful in polarization-independent waveguides relying full bandgaps\cite{19}, enhanced nonlinear optical effects\cite{20}, and polarization division multiplexing\cite{21}. Topologically protected polarization-independent optics would give them additional resistance to perturbation. It is worth noting that dual-polarization second-order photonic topological states have been proposed by Chen et al. recently, based on a topologically optimized geometric structure within a square-lattice\cite{22}. However, eigenfrequencies of topological states for the two polarizations are not the same, despite they have a common topological bandgap.

In this paper, a 2D second-order photonic topological insulator (SPTI) is proposed, of which the topological states are polarization-independent. The square-lattice PC having a fishnet structure is made by an elliptic metamaterial. The permittivity is anisotropic and nevertheless, the geometry structure is rather simple compared with the previously proposed topologically optimized structure. That the CBG is either trivial or topological depends on the choice of the unit cell (UC) for both TM and TE modes. The proposed SPTI can host topological edge states and corner states for the two modes at the same time. Our results show polarization-independent topological corner states based a SPTI is not guaranteed by a common topological bandgap. However, we find that locations of bandgaps and corner states in the frequency spectrum can be manipulated independently by the out-of-plane permittivity $\varepsilon_\perp$ and in-plane permittivity $\varepsilon_\varparallel$ for TM and TE modes, respectively, which gives an effective way to achieve overlapped corner states for the two modes. On this basis, corner states independent of polarization can be realized by choosing appropriate $\varepsilon_\perp$ and $\varepsilon_\varparallel$. Numerical simulations further show the corner states are topologically protected, with strong robustness to disorders and defects. Our work shows potential applications in polarization-independent topological photonic devices.

\section{structure design and band topology}
\begin{figure}[htbp]
\centering\includegraphics[width=11cm]{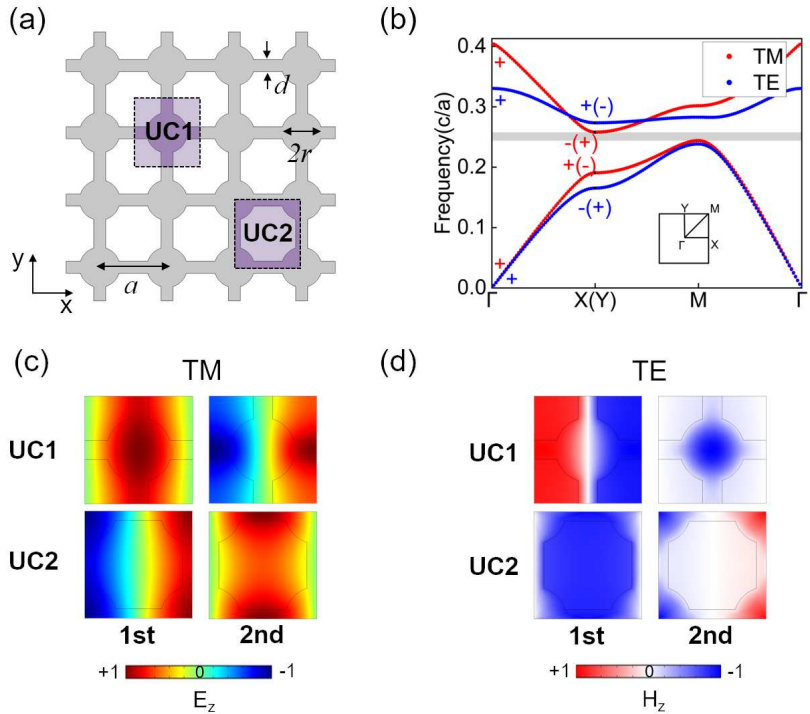}
\caption{\label{fig_1}(a) Fishnet PC and two kinds of unit cells (UC), UC1 and UC2. (b) Band structures of TM and TE modes, denoted by red and blue dot-lines, respectively. Even and odd parities of UC1 (UC2) at high symmetric points are indicated by plus and minus symbols, colored in red and blue for TM and TE modes, respectively. (c) $E_z$ field patterns of the two TM bands at the X point for UC1 and UC2. (d) $H_z$ field patterns of the two TE bands at the X point for UC1 and UC2.}
\end{figure}

For PCs, it is well known that TM bandgaps are favored in dielectric rods, while TE bandgaps prefer dielectric veins\cite{add1}. From this, the proposed square-lattice PC is constructed by thin dielectric veins with dielectric rods located at lattice sites, as shown in Fig.~\ref{fig_1}(a). $a$ is the lattice constant, and the circle radius $r$ and vein width $d$ are 0.3$a$ and 0.18$a$, respectively. The dielectric material is anisotropic, an elliptic metamaterial with the permittivity $\varepsilon=(\varepsilon_\varparallel,\varepsilon_\varparallel,\varepsilon_\perp)=(16.9,16.9,10)$. Generally, topological corner states lie in a topological bandgap\cite{13}, and hence a topological CBG of TM and TE polarizations is the prerequisite for polarization-independent topological corner states. The choice of the elliptic metamaterial is based on the consideration that bandgap locations of TM and TE polarizations in the frequency spectrum can be manipulated independently by $\varepsilon_\perp$ and $\varepsilon_\varparallel$, respectively. In practical, we can use the multilayer model to construct the anisotropic permittivity\cite{23}. The multilayer consists of two alternative dielectrics with high and low permittivity, and it is placed horizontally in the x-y plane. According to the formulisms (16) and (17) proposed in ref\cite{add0}, the PC slab with permittivity (16.9,16.9,10) can be approximately built by the high dielectric with the permittivity of 17.67 and the air layer when the filling ratio of high dielectric is 0.954. Herein, the calculation of band structures and numerical simulations are based on the finite element method using the commercial software COMSOL Multiphysics.

Based on the common square lattice, two kinds of unit cells (UCs), UC1 and UC2, are selected in Fig.~\ref{fig_1}(a). Note that the two UCs are consistent with each other after shifting the center of one of the UCs by half of the period along $x$ and $y$ directions. Therefore, they share the same band structure as plotted in Fig.~\ref{fig_1}(b), with red and blue dot-lines denoting the TM and TE modes, respectively. One can find that there is a CBG indicated by the gray region lying between the first and second bands of TM modes. However, for the two UCs, the CBG possesses different topological behaviors characterized by the 2D Zak phase [see Appendix A], which has the following form\cite{24,25,26}:

\begin{equation}
\theta_j^{Zak}=\int dk_xdk_y\text{Tr}[\hat{A_j}(k_x,k_y)],
\end{equation}
where $j=x$ or $y$, and the Berry connection $\hat{A_j}=i\langle u(\textbf{k})|\nabla_{k_j}| u(\textbf{k})\rangle$ with $u(\textbf{k})$ being the \textcolor{red}{periodic part of} the Bloch function. The 2D Zak phase can also be understood by the 2D bulk polarization via $\theta_j^{Zak}=2\pi P_j$ with

\begin{equation}
P_j=\frac{1}{2}(\sum_n q_j^n\;\text{mod 2})\text{,}\qquad (-1)^{q_j^n}=\frac{\eta(X_j)}{\eta(\Gamma)}
\end{equation}
where $P_j$ is determined by the parity $\eta$ associated with $\pi$ rotation at $\Gamma$ and X(Y) points and the summation is over all the occupied bands below the bandgap. Here, $P_x$ is equal to $P_y$, namely, $P_x=P_y$, due to the $C_4$ symmetry\cite{27,28}. Eigenfield patterns at the X point of the two bands for TM and TE modes are shown in Figs.~\ref{fig_1}(c) and \ref{fig_1}(d), respectively, with the monopole an even parity and dipole an odd parity. As can be seen, the parities of the two bands at the X point have an inversion between UC1 and UC2 for both the two modes, whereas the parities at the $\Gamma$ point stay the same. Moreover, parities of the same UC at the X point are opposite for TM and TE modes, which gives the same UC distinct topological properties for the two modes. Concretely, for TM modes, the distinct parties of UC1 at the X and $\Gamma$ points give the 2D bulk polarization $(P_x,P_y)$ a value of $(0,0)$ and the 2D Zak phase $(\theta_x^{Zak},\theta_y^{Zak})$ a value of $(0,0)$, while the same parity of UC2 at the X and T points makes $(P_x,P_y)=(\frac{1}{2},\frac{1}{2})$ and $(\theta_x^{Zak},\theta_y^{Zak})=(\pi,\pi)$. The opposite is true for the TE modes. As a result, the bandgap of UC1 is trivial and of UC2 is topological for TM modes, and it is reversed for TE modes.

\begin{figure*}[ht!]
\centering\includegraphics[width=13cm]{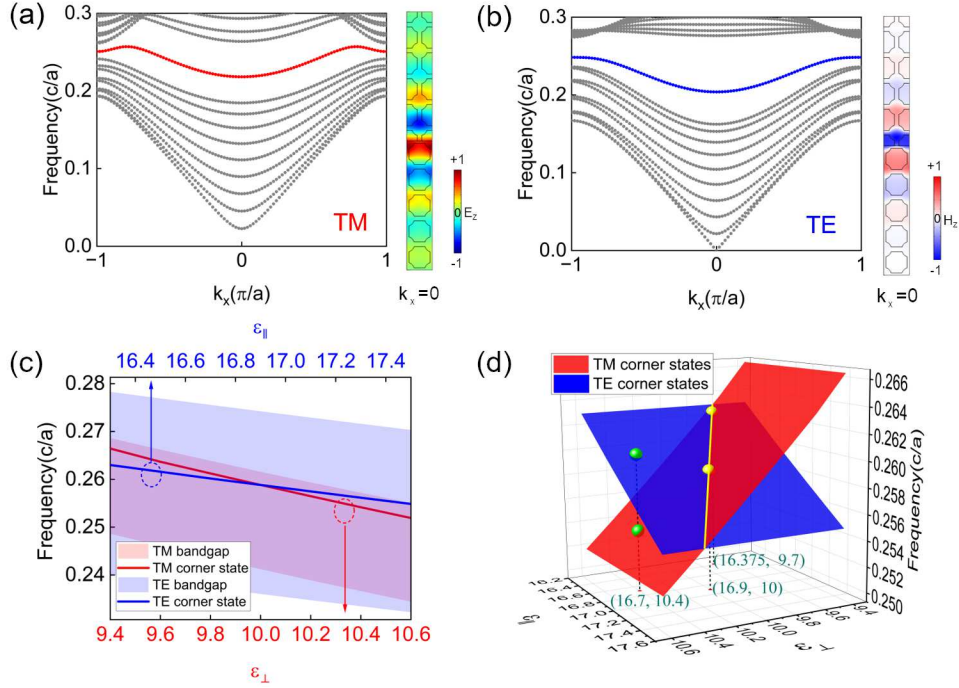}
\caption{\label{fig_2}Projected band structures of (a) TM and (b) TE modes, with edge modes colored in red and blue, respectively. Eigenfields at $k_x=0$ show the edge modes can be well confined at the interface between UC1s and UC2s for both TM and TE modes. (c) Dependence of bandgaps and eigenfrequencies of one of the corner states on $\varepsilon_\perp$ and $\varepsilon_\varparallel$ for the two modes. The area shaded in light red indicates TM bandgaps, while the area shaded in light blue indicates TE bandgaps. The red and blue lines denote one of the corner states of TM and TE modes, respectively. (d) TM corner states (colored in red) and TE corner states (colored in blue) under any combination of $\varepsilon_\perp$ and $\varepsilon_\varparallel$ in the same parameter range of (c). The yellow intersecting line denotes the combinations that have overlapped corner states. The yellow points on the intersecting line is two of the combinations, and their anisotropic permittivity $(\varepsilon_\varparallel,\varepsilon_\varparallel,\varepsilon_\perp)$ are (16.9,16.9,10) and (16.375,16.375,9.7), respectively. The green points are the two points that share the same anisotropic permittivity (16.7,16.7,10.4) but have different eigenfrequencies.}
\end{figure*}

\begin{figure*}[ht!]
\centering\includegraphics[width=12.8cm]{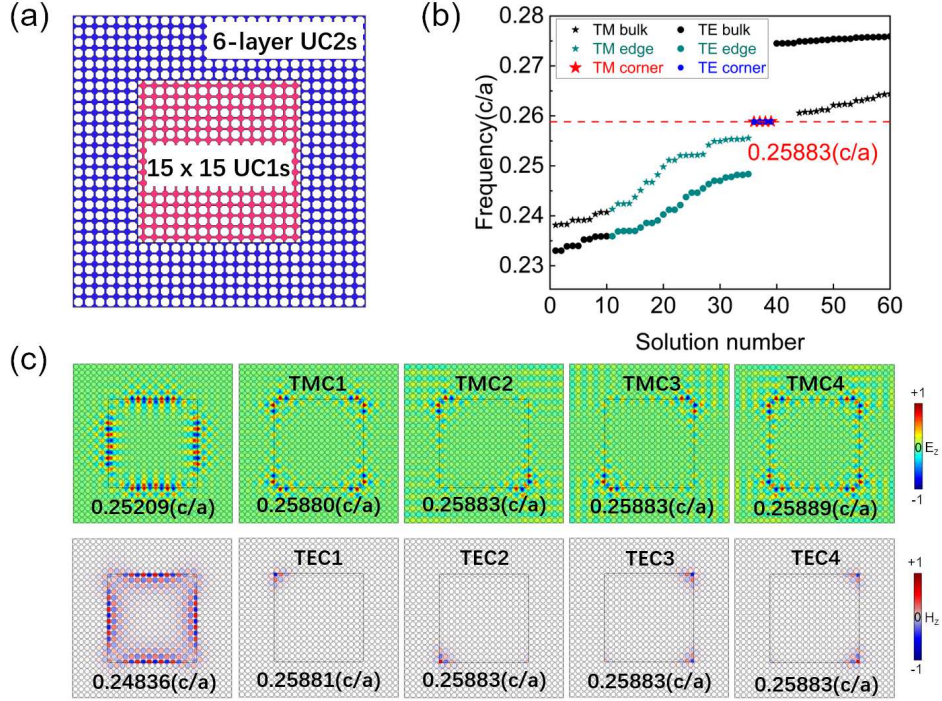}
\caption{\label{fig_3}(a) Schematic of the finite-size box-shaped PC, with 15$\times$15 UC1s surrounded by 6-layer UC2s. (b) Eigenfrequencies of the box-shaped PC. TM and TE modes are denoted by pentagons and circles, with their corner states colored in red and blue, respectively. Edge modes are shown as cyan. (c) Eigenfields of the overlapped edge and corner modes.}
\end{figure*}

\begin{figure*}[ht!]
\centering\includegraphics[width=15.5cm]{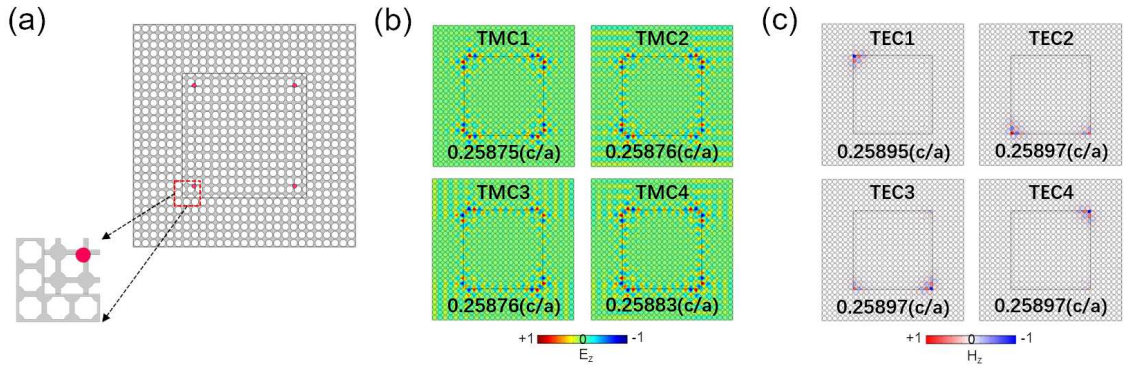}
\caption{\label{fig_4}(a) Box-shaped PC with four disorders (red dots) around four corners of the internal PC composed UC1s. The enlarged view shows one of the four disorders, with $10\%$ decrease in radius and $0.1a$ deviation from the lattice site along $x$ and $y$ directions. Eigenfields of four corner modes of (b) TM and (c) TE modes, under the influence of the disorders.}
\end{figure*}

\begin{figure*}[ht!]
\centering\includegraphics[width=15.5cm]{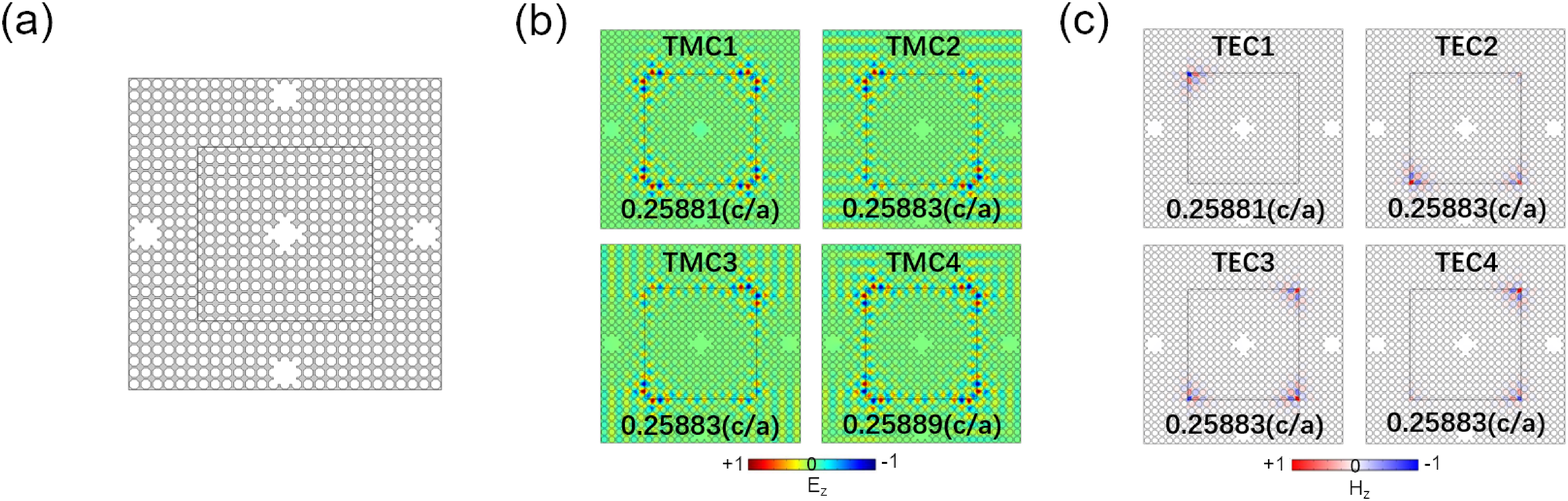}
\caption{\label{fig_5}(a) Box-shaped PC with defects produced by removing five UC1s in the center and four UC2s near the edge of the PC. Eigenfields of four corner modes of (b) TM and (c) TE modes, under the influence of the defects.}
\end{figure*}
 
\section{polarization-independent topological corner states}
The topological distinction between UC1 and UC2 ensures the existence of topological edge states\cite{29,30,31,32}. To show this, we construct a supercell composed of five UC1s and five UC2s along the $y$ direction, and projected band structures are shown in Figs.~\ref{fig_2}(a) and \ref{fig_2}(b) for TM and TE modes, respectively. In the calculation, periodic boundary conditions are applied to the $x$ direction. As can be seen, there is one in-gap edge state for both the two modes, which does not occupy the whole bulk bandgap and canbe confined at the interface between UC1s and UC2s. 
Since therer is a $C_4$ symmetry for the PC, we can define a corner topological index: $Q^c=\frac{1}{4}([X_1]+2[M_1]+3[M_2])$, where $[\Pi_p]=\#\Pi_p-\#\Gamma_p$ and $\#\Pi_p$ is defined as the number of bands below the bandgap with rotation eigenvalues $\Pi_p^n=e^{[2\pi i(p-1)/n]}$ for p=1, 2, 3, 4. For the nontrivial TM and TE cases, they both have $[X_1]=-1$, $[M_1]=-1$, $[M_2]=0$. Therefore, the corner topological index is $Q^c=\frac{1}{4}$ for both the two modes, indicating $\frac{1}{4}$ fractionalized corner states at each of the four corners\cite{32}.  
It is noteworthy that the existence of polarization-independent corner states is not guaranteed by the CBG. In Fig.~\ref{fig_2}(c), we change $\varepsilon_\perp$ and $\varepsilon_\varparallel$ in the certain range near $(16.9,16.9,10)$ to solely adjust the positions of supercell bandgaps in the frequency spectrum for TM and TE modes, respectively. Specifically, for the TM band gap, we increase $\varepsilon_\perp$ from 9.4 to 10.6 and keep $\varepsilon_\varparallel$ at any value, while for the TE band gap, we increase $\varepsilon_\varparallel$ from 16.3 to 17.5 and keep $\varepsilon_\perp$ at an arbitrary value. As can be seen, the positions of the two bandgaps descend as the corresponding permittivity increases, and the TM bandgap (light red area) is completely embedded in the TE bandgap (light blue area), forming the CBG. We also calculate the eigenfrequencies of TM (red line) and TE (blue line) corner states from the finite-size box-shaped PC shown in Fig.~\ref{fig_3}(a), and find that they are in the CBG and the variation trend of the corner states with the permittivity is the same as that of the bandgaps. Since the two kinds of polarized corner states are independent of each other, in order to search for the overlapped ones, we plot their eigenfrequencies under any combination of $\varepsilon_\perp$ and $\varepsilon_\varparallel$ in Fig.~\ref{fig_2}(d). It can be observed that corner states of the two modes do not coincide with each other except on the yellow intersecting line. The yellow points on the intersecting line are two of the combinations that have the overlapped corner states, and the corresponding anisotropic permittivities $(\varepsilon_\varparallel,\varepsilon_\varparallel,\varepsilon_\perp)$ are (16.9,16.9,10) and (16.375,16.375,9.7). As a contrast, green points are the two points that share the same anisotropic permittivity (16.7,16.7,10.4) but have different eigenfrequencies. Therefore, the anisotropic permittivity provides an additional freedom to manipulate the location of corner states of the two modes, making the corner states either polarization-independent or polarization-separable [see Appendix B].

To verify the existence of the polarization-independent corner states, a box-shaped PC of finite size is constructed, which is composed of $15\times15$ UC1s surrounded by six-layer UC2s, as shown in Fig.~\ref{fig_3}(a). The calculated eigenfrequencies of TM and TE modes based on the anisotropic permittivity (16.9,16.9,10) are shown in Fig.~\ref{fig_3}(b). As can be seen, both of them show gapped edge modes and four in-gap corner modes. Red and blue dotted lines go through the overlapped corner and edge states, respectively. In Fig.~\ref{fig_3}(c), eigenfields of these topological states indicate that the edge modes can be well confined along the whole interface between UC1s and UC2s, while the corner states are highly localized at the corners of the internal PC formed by the UC1s. Remarkably, topological corner states for the two modes do share the same eigenfrequencies, and the common eigenfrequency of the corner states is 0.25883(c/a). This is different from the previously reported dual-polarization topological corner states, which possess the topological CBG, but their eigenfrequencies are not overlapped at all\cite{22}.

\begin{figure*}[ht!]
\centering\includegraphics[width=16cm]{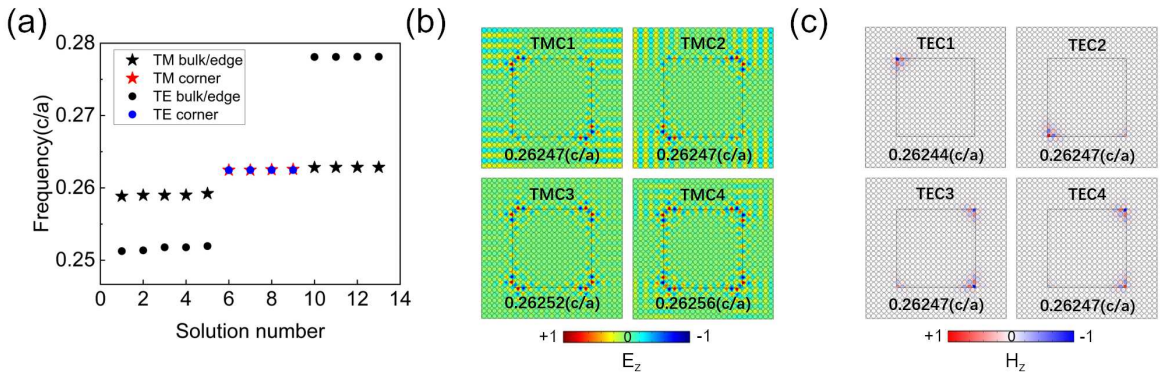}
\caption{\label{fig_6}(a) Eigenfrequencies of the box-shaped PC with an anisotropic permittivity of (16.375,16.375,9.7), showing overlapped corner states of TM and TE modes. Pentagons and circles denote TM and TE modes, and their corner states are colored in red and blue, respectively. (b). Eigenfields of the corner states of TM modes. (c) Eigenfields of the corner states of TE modes.}
\end{figure*}

\begin{figure*}[ht!]
\centering\includegraphics[width=16cm]{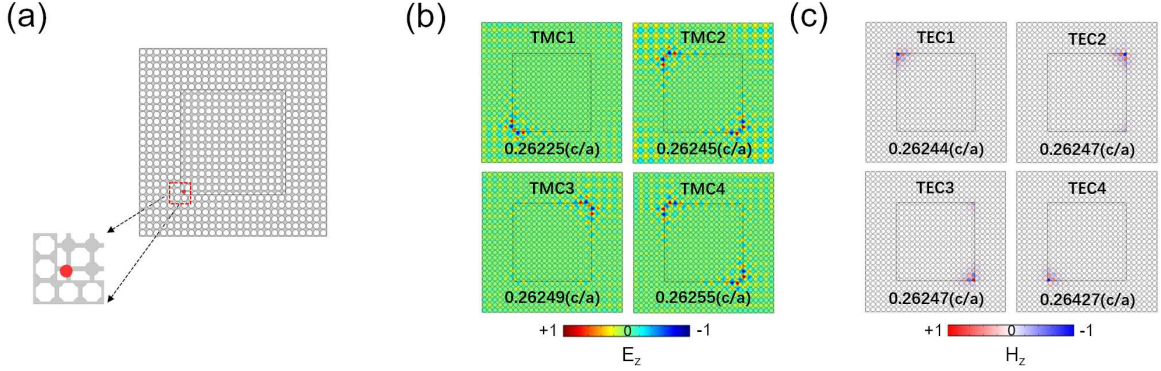}
\caption{\label{fig_7}(a) Box-shaped PC with a disorder (red dot) located at the left bottom corner of the internal PC composed UC1s. The enlarged view shows the single disorder, with $10\%$ decrease in radius and $0.1a$ deviation from the lattice site along $x$ and $y$ directions. Eigenfields of four corner modes of (b) TM and (c) TE modes, under the influence of the disorder.}
\end{figure*}

\begin{figure*}[ht!]
\centering\includegraphics[width=16cm]{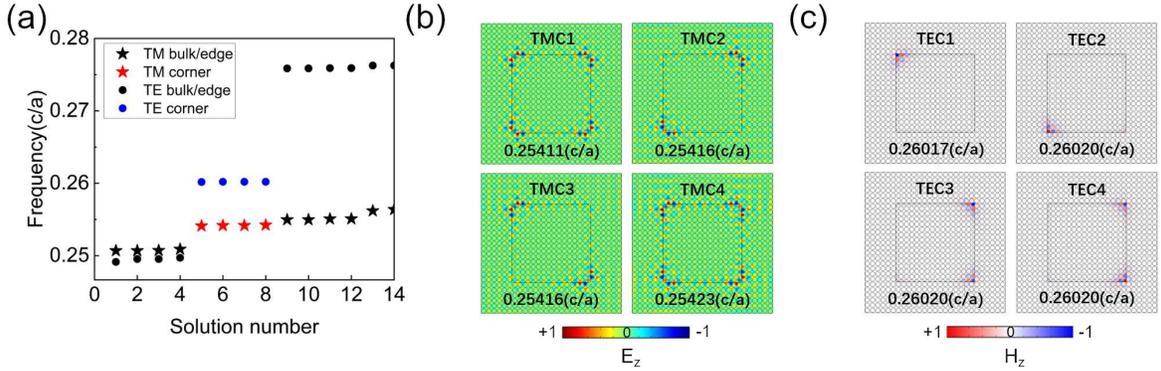}
\caption{\label{fig_8}(a) Eigenfrequencies of the box-shaped PC with an anisotropic permittivity of (16.7,16.7,10.4), which shows corner states of TM and TE modes are not overlapped. Pentagons and circles denote TM and TE modes, and their corner states are colored in red and blue, respectively. (b). Eigenfields of the corner states of TM modes. (c) Eigenfields of the corner states of TE modes.}
\end{figure*}

The polarization-independent photonic corner states are topologically protected due to their topology origin\cite{36,37}. To verify this, we introduce four disorders marked by red dots around the four corners into the instead perfect PC, as shown in Fig.~\ref{fig_4}(a). The enlarged view in Fig.~\ref{fig_4}(a) exhibits the single disorder, with $10\%$ decrease in radius and $0.1a$ deviation from the lattice site along $x$ and $y$ directions. Eigenfields of the corner states of TM and TE modes are shown in Figs.~\ref{fig_4}(b) and \ref{fig_4}(c), respectively, from which we can see that the corner states still exist with negligible offsets of the eigenfrequencies. Beyond that, defects, produced by removing five UC1s and four UC2s in the center and near the edge of the PC respectively, are also introduced, as shown in Fig.~\ref{fig_5}(a). As can be seen in Figs.~\ref{fig_5}(b) and \ref{fig_5}(c), since the defects are far away from the corners, the eigenfrequencies of the corner states for the TM and TE modes remain unchanged, although the defects have a more destructive effect on the PC structure\cite{34}.

\section{conclusion}
In summary, a polarization-independent SPTI is achieved, based on a 2D square-lattice PC. The dielectric is an elliptic metamaterial, and the geometric structure is rather simple nevertheless. By selecting appropriate geometric parameters and anisotropic permittivity, a CBG is can be obtained for TM and TE modes. That the CBG of a certain UC is either trivial or topological depends on the polarization modes. Topological corner states of TM and TE modes can coexist in the CBG, but only the combinations of in-plane permittivity $\varepsilon_\varparallel$ and out-of-plane permittivity $\varepsilon_\perp$ that lie on the intersecting line in the eigenfrequency-permittivity space can make them overlapped. Numerical simulations further show they have strong robustness to disorders and defects. The proposed scheme can also be extended to corner states induced by the quadrupole topological phase in square-lattices, pseudo-spin and valley-spin degrees of freedom. Our work would pave the way toward designing high-performance polarization-independent topological photonic devices, such as the polarization-independent topological laser and coupled cavity-waveguide. 

\begin{acknowledgments}
The work was jointly supported by the National Natural Science Foundation of China (12064025, 12264028) and Natural Science Foundation of Jiangxi Province (20212ACB202006) and Major Discipline Academic and Technical Leaders Training Program of Jiangxi Province (20204BCJ22012).
\end{acknowledgments}

\appendix
\section{Tight-binding model}
The tight-binding model gives the topological phase transition between the UC1 and UC2 a well description, in which one can take the dielectric rods as lattice sites for TM modes while the air holes act the part for TE modes. The Hamiltonain has the following form,

\begin{equation}
H=-\sum_{ij} t_{ij}c_i^{\dag}c_j,
\end{equation} 
where $t_{ij}$ is the hopping amplitude between the nearest lattice sites and $c_i^{\dag}(c_j)$ is the creation (annihilation) operator. As there is only one lattice cite in UCs, the one band below the photonic bandgap can be expressed as

\begin{equation}
E=-t_0(e^{ik_x}+e^{-ik_x}+e^{ik_y}+e^{-ik_y})
=-2t_0(\cos k_x+\cos k_y). 
\end{equation}

Look at TM modes first, for UC1, the lattice site choosed as the inversion center is at the center of the UC1, and the inversion operator is $I=1$. Hence, parities at $\Gamma$ and X points are the same. For UC2, lattice sites are at the four corners and the inversion operator $I=e^{\pm i(k_x+k_y)}$ hinges on which lattice site is referenced. Thus, the parity is +1 at the $\Gamma=(0,0)$ point, while it is -1 at the $X=(\pi,0)$ point\cite{26}.

For TE modes, lattice sites of UC1 choosed as the inversion center are at the four corners, since the air holes instead of the dielectric rods act the role of lattice sites. For UC2, the lattice site choosed as the inversion center is at the center of UC2. As a consequence, parities at $\Gamma$ and X points are oppostie for UC1, while they are the same for the UC2. The results are consistence with parities showed in Fig.~\ref{fig_1}(b).

\section{Switch between polarization-independent and polarization-separable corner states}
Here, we would like to show another anisotropic permittivity lying on the intersecting line that can achieve polarization-independent topological corner states. The anisotropic permittivity is $(16.375,16.375,9.7)$, as indicated in the Fig.~\ref{fig_2}(e). Fig.~\ref{fig_6}(a) shows the calculated eigenfrequencies, from which we can see that the corner states of the two modes can be overlapped under this permittivity. Eigenfrequencies and eigenfields of the corner states are shown in Figs.~\ref{fig_6}(b) and \ref{fig_6}(c), and one can see the overlapped eigenfrequency is 0.26247(c/a). In Fig.~\ref{fig_7}, if we introduce a single disorder into the box-shaped PC, the corner states still survive with litte frequency shit, but monopole and quadrupole of TM modes no longer exist due to the broken of the $C_4$ symmetry of the box-shaped PC.  

Noting that if the anisotropic permittivity is off the intersecting line, polarization-independent corner states will be changed into polarization-separable corner states. As shown in the Fig.~\ref{fig_2}(e), if the anisotropic permittivity is (16.7,16.7,10.4), eigenfrequencies of the corner states of the two modes are apart from each other. In detail, we plot the eigenfrequencies in Fig.~\ref{fig_8}(a), and one can find that none of the four corner states of the two modes are the same, and the maximum frequency difference between the two modes is 0.00606(c/a) as shown in  Figs.~\ref{fig_8}(b) and \ref{fig_8}(c).

\bibliography{refs}

\end{document}